\documentclass[prd,nofootinbib,12pt]{revtex4}
\usepackage{amsmath}

\begin{document}

\def\beq{\begin{equation}}
\def\eeq{\end{equation}}
\def\ber{\begin{eqnarray}}
\def\eer{\end{eqnarray}}
\def\apj{{Astrophys.\@ J.\ }}
\def\mn{{Mon.\@ Not.@ Roy.\@ Ast.\@ Soc.\ }}
\def\asta{{Astron.\@ Astrophys.\ }}
\def\aj{{Astron.\@ J.\ }}
\def\prl{{Phys.\@ Rev.\@ Lett.\ }}
\def\prd{{Phys.\@ Rev.\@ D\ }}
\def\pd{{Phys.\@ Rev.\@ D\ }}
\def\nucp{{Nucl.\@ Phys.\ }}
\def\nat{{Nature\ }}
\def\plb {{Phys.\@ Lett.\@ B\ }}
\def \jetpl {JETP Lett.\ }
\def\etal{{\it et al.}}
\def\ie {{\it ie~}}
\def\half{{1\over 2}}

\title{APSIS --- an Artificial Planetary System in Space to probe
extra-dimensional gravity and MOND}

\author{Varun Sahni}\email{varun@iucaa.ernet.in}
\affiliation{Inter-University Centre for Astronomy and Astrophysics, Post Bag
4, Ganeshkhind, Pune 411~007, India}
\author{Yuri Shtanov}\email{shtanov@bitp.kiev.ua}
\affiliation{Bogolyubov Institute for Theoretical Physics, Kiev 03680, Ukraine}

\begin{abstract}
A proposal is made to test Newton's inverse-square law using the perihelion
shift of test masses (planets) in free fall within a spacecraft located at the
Earth--Sun L2 point. Such an {\em Artificial Planetary System In Space} (APSIS) will
operate in a drag-free environment with controlled experimental conditions and
minimal interference from terrestrial sources of contamination. We demonstrate
that such a space experiment can probe the presence of a `hidden' fifth
dimension on the scale of a micron, if the perihelion shift of a `planet' can
be measured to sub-arc-second accuracy. Some suggestions for spacecraft design
are made.
\end{abstract}

\maketitle

\section{Introduction}

Cosmology at the turn of this century appears to stand at new crossroads.

Remarkably precise recent observations have, one the one hand, confirmed long
standing theoretical ideas (inflation, baryon oscillations) and, on the other
hand, provided glimpses into unexpected territories and landscapes (dark matter
and dark energy). One of the central issues facing the cosmologist in the new
century is an explanation for the multifarous properties of an accelerating
universe and the plausible existence of extra dimensions. Indeed, it now
appears \cite{sn,eisenstein05,astier05} that only about 4\% of the matter
content of the universe is baryonic in form. To account for the remaining 96\%,
one usually invokes the presence of `dark matter' ($\sim 26\%$) and `dark
energy' ($\sim 70\%$). While dark matter successfully explains several
different sets of observations, its present avatar --- cold dark matter
--- is currently facing an increasing number of observational challenges
including galaxy cores which appear to be shallower than the cuspy cores
predicted by CDM, and an over-abundance of dwarf galaxies predicted by this
scenario to exist both in voids and in our local group, and not seen in either.
While it may be that traditional remedies to these problems (baryonic feedback,
making the dark matter `warm' instead of `cold', using a scalar field to
describe dark matter, etc.\@) may alleviate some of the tension between theory
and observations
\cite{sahni_greece,abazajian06,sahni_wang,um00,hu_barkana-gruz}, it could also
be that the current situation warrants a more fundamental revision of our
understanding of the basic laws governing gravity. An example of a radical
approach which attempts the latter is MOdified Newtonian Dynamics (MOND), a
phenomenological model originally suggested in 1983, which gives impressive
results in explaining the flat rotation curves of galaxies and some other
observations \cite{milgrom,sanders}.

In contrast to dark matter, which is assumed to be pressureless and prone to
gravitational clustering, dark energy is virtually unclustered and endowed with
a large negative pressure which allows it to explain the current acceleration
of the universe. Like it is the case with dark matter, the existence of dark
energy is largely hypothetical, its {\em raison d'\^{e}tre\/} being
observations of cosmic acceleration, which are not easily explained by a more
conventional matter source. However, the unevolving nature of the simplest dark
energy candidate --- the cosmological constant --- implies that the ratio of
the density in the latter ($\rho_\Lambda \simeq 10^{-47}$~GeV${}^4$) to the
radiation/matter density is an increasingly small number at early times. For
instance, $\rho_\Lambda \simeq 10^{-123} \rho_P$ at the Planck time $\sim
10^{-43}$ seconds after the Big Bang (here, $\rho_P$ is the Planck density, the
only natural value at that time). This gives rise (according to one's
perspective) either to an initial `fine-tuning' problem or to a `cosmic
coincidence' conundrum \cite{DE_review1}.

Keeping these issues in mind, the prevailing views on dark matter and dark
energy have, in recent years, been supplemented by new ideas, which see the
recent observations as lending support to the possibility that our traditional
theories of gravity may need reformulation either in regions of small
acceleration (MOND) or on very large scales (braneworld models \cite{brane},
modified gravity theories \cite{DE_review2}, etc).\footnote{For other
alternative explanations of dark matter and dark energy see
\cite{alternatives1,alternatives2}}

For instance, MOND assumes that Newton's law of inertia ($F = ma$) is modified
at sufficiently low accelerations ($a < a_0$), so that
  \beq
   {\bf F} = m{\bf a} \mu \left(\frac{a}{a_0} \right) \, ,
    \eeq
where $\mu(x) = x$ when $x \ll 1$, and $\mu(x) = 1$ when $x \gg 1$
\cite{milgrom,sanders}. It is easy to see that this leads to the following
limiting velocity of a body orbiting a point mass $M$:
\beq v^4 = GMa_0 \, .
\eeq
In other words, for sufficiently low values of acceleration, this theory
predicts flat rotation curves (which are formally infinite in extent).
Surprisingly, the value needed to explain observations is $a_0 \sim
10^{-8}$~cm/s$^2$, which is of the same order as $cH_0$\,! The increasing
success rate of MOND in explaining observational data has led to an appreciable
growth in the number of recent research publications on this subject. Clearly
of interest would be tests which might simulate MOND-like conditions in the
controllable environment of a laboratory. One such space experiment will be
discussed later in this paper.

We end this introduction by noting a strange coincidence: the MOND acceleration
$a_0$ is tantalizingly close to that associated with the Pioneer anomaly
\cite{pioneer}. This is the anomalous acceleration experienced by the
spacecraft Pioneer 10 and 11, which was noticed in 1980 after Pioneer 10 had
passed a distance of $\sim 20$ astronomical units from the Sun. (Pioneer 10 has
now left the solar system.) The acceleration is directed towards the Sun and
has the value \cite{pioneer} $(8.60 \pm 1.34)\times 10^{-8}$~cm/s$^2$. Efforts
to find a conventional explanation for this effect in terms of spacecraft
design, leakage, and the influence of the solar wind have so far proved elusive
and proposals have been made, both by NASA and the European Space Agency, for a
dedicated space mission to probe the Pioneer anomaly \cite{pioneer1,nasa}.

\section{Extra dimensions}

The possibility that space could have more than three dimensions was originally
suggested in the seminal works of Kaluza (1921) and Klein (1926), who
demonstrated that a compact (circle-like) fifth dimension would unify  gravity
with the electromagnetic force \cite{kk}. The Kaluza--Klein program was pursued
with great enthusiasm during the 1980's, the main objective then being the
recovery of gauge fields and symmetries of the standard model from compact
(hidden) dimensions. However, since the size of the extra dimensions was close
to the Planck scale, ${\cal R} \sim \ell_P \simeq 10^{-33}$~cm, direct
observational evidence of these dimensions was virtually impossible, and extra
dimensions in such theories were kept well hidden. A paradigm shift in our
perception of a multi-dimensional cosmos occurred when it was suggested that
extra dimensions, though compact, may be much larger than the Planck size
\cite{antoniadis}, and even macroscopic \cite{dvali}, ${\cal R} \lesssim 1$~mm.
The rationale for such an approach was the long-standing hierarchy problem in
physics which arises because the Planck scale is so much higher than other mass
scales in particle physics (for instance $M_P/M_W \sim 10^{17}$, where $M_W$ is
the mass of the vector bosons which mediate the weak force). Within this new
higher-dimensional framework, the fundamental scale of gravity can be much
lower than $M_P$, and a simple example shows how this can be achieved.

Consider two test masses $m_1$ and $m_2$ separated by a distance $r \ll {\cal
R}$ in a $(4+n)$-dimensional universe and interacting via the gravitational
potential\footnote{The discussion here closely follows that in
\cite{dvali,ss02a}.}
\begin{equation} \label{pot1}
V(r) \sim \frac{m_1\, m_2}{M^{n+2}}\, \frac{1}{r^{n+1}}\, , \quad r \ll {\cal
R},
\end{equation}
where $M$ is the $(4+n)$-dimensional Planck mass. If the same two particles are
placed much further apart, then, because the gravitational field lines
associated with $m_1$ and $m_2$ do not have room to propagate in the extra
dimensions at such large distances, the potential at large separations becomes
\begin{equation} \label{pot2}
V(r) \sim \frac{m_1\, m_2}{M^{n+2} {\cal R}^n}\, \frac{1}{r} \sim \frac{m_1\,
m_2}{M_P^2}\, \frac{1}{r} \, , \quad r \gg {\cal R} \, .
\end{equation}
From (\ref{pot1}) \& (\ref{pot2}) one finds that the effective four-dimensional
Planck mass is simply given by $M^{2}_P \sim M^{n+2}_N {\cal R}^n$. From
(\ref{pot1}) we find that gravity becomes higher-dimensional on length scales
smaller than ${\cal R}$; substituting ${\cal R} \sim 1$~mm and $n = 2$, one
gets $M \sim 1$~TeV. The theory of macroscopic compact extra dimensions can be
tested by traditional `table top' experiments which probe the inverse-square
law on scales down to $\sim 0.1$~mm \cite{hoyle01}. Other tests include
sufficiently energetic collisions on the LHC, NLC etc.\@ \cite{dvali}.

Of course, the scheme described above could not be realized in the original
Kaluza--Klein approach since the presence of extra dimension of millimeter size
would be well observable in ordinary, non-gravitational, physics.  Therefore,
an interesting alternative approach adopted in \cite{dvali} was connected with
the braneworld concept. In this scenario, our (3+1)-dimensional universe is
thought to be a brane (from the word `membrane') embedded in a
higher-dimensional `bulk' space time with large extra spatial dimensions. The
fields of Standard Model are confined to move along the brane and, therefore,
do not `feel' the presence of extra dimensions, whereas gravity can propagate
in the bulk and have the properties described above.\footnote{In an earlier
proposal of this kind \cite{antoniadis}, the matter fields of the Standard
Model were localized at the orbifold fixed points while the gauge fields
propagated also in extra dimensions of TeV size.}

The next fruitful idea was to consider bulk spaces with {\em
infinite\/} noncompact extra dimensions \cite{rs99,rubsh,akama,rub_uspekhi}. In
this paradigm, gravity on the brane remains four-dimensional on sufficiently
large scales due to the effect of curvature of the bulk space. A seminal model
of this kind was put forward by Randall and Sundrum \cite{rs99}.  It has one
infinite extra dimension, and the space-time metric in this model has the
form\footnote{In the Randall-- Sundrum model the small value of the true
five-dimensional Planck mass is related to its large effective four-dimensional
value by the extremely strong warp of the five-dimensional space.}
\beq
  ds^2 = e^{-2k|y|} \eta_{\mu\nu} dx^\mu dx^\nu - dy^2 ~.
\eeq
Writing the full gravitational potential between two point masses on the brane
in this model as $V(r) = V_0(r) \left[1 + \Delta (r)\right]$, one can calculate
the series expansion for the correction term $\Delta (r)$ in the form
\cite{callin_ravndall}
\ber k r \ll 1: \quad
\Delta &=& \frac{4}{3\pi k r} - \frac{1}{3} - \frac{1}{2\pi} k r \ln k r +
0.089237810\, k r + {\cal O}\left[(k r)^2 \right] \, ,\\
k r \gg 1: \quad \Delta &=& \frac{2}{3 (k r)^2} - \frac{4\ln k r}{(k r)^4} +
\frac{16 - 12\ln 2}{3(k r)^4} + {\cal O} \!\left[ \frac{(\ln k r)^2}{(k r)^6}
\right] \! .
\eer
Thus, on length scales $r$ smaller than the curvature radius $k^{-1}$ of the
fifth dimension, gravity becomes five-dimensional, and the gravitational
potential changes from its familiar four-dimensional form $V(r) \propto 1/r$ to
the five-dimensional $V(r) \propto 1/r^2$.  On scales larger than the curvature
radius, the leading correction to the potential is $V(r) \propto 1/r^3$.

Considerable support for the multi-dimensional viewpoint
comes from string and M-theory, in which extra dimensions play a crucial role
in the unification of all forces at a fundamental level.  In fact, it was the
famous supergravity model of Ho\v{r}ava and Witten \cite{hwitten}, representing
two ten-dimensional branes connected by large eleventh dimension, that inspired
the model by Randall and Sundrum \cite{rs99}.

It is interesting to note that, while higher-dimensional theories generically
predict departures from the inverse-square law on small scales
\cite{maartens,bronnikov06}, such departures can also arise in other theories.
For instance, models which modify the general-relativistic Einstein--Hilbert
action can lead to MOND-type effects at low accelerations and/or departures
from Newton's law on small scales (see, for instance, \cite{navarro05}).

It is important to add that extra-dimensional models as well as
four-dimensional models which modify the general-relativistic Einstein--Hilbert
action, have several important cosmological properties. Amongst these is the
attractive possibility of explaining cosmic acceleration without dark energy
and of describing both early and late-time acceleration within a single unified
setting (see \cite{sahni_kyoto,modified-grav} and references therein).

Having briefly discussed the theoretical motivation for expecting departures
from Newton's laws on small scales and/or at low accelerations, we now outline
a space experiment which attempts to detect these new phenomena.

\section{An Artificial Planetary System In Space (APSIS)}

\subsection{The perihelion shift}

As briefly discussed above, several alternative theories of gravity predict a
force law which differs from Newton's inverse-square law on small scales. This
departure usually occurs in one of two ways:
\begin{enumerate}

\item Due to the presence of a Yukawa-like force on small scales
\beq
V(r) = - \frac{G_\infty m_1m_2}{r} \left [ 1 \pm \alpha \exp{(-r/r_0)}\right]
\label{eq:yukawa}
\eeq
which falls off exponentially with increasing distance from the source.  Here,
$G_\infty$ is the value of Newton's gravitational constant (formally) measured
at infinity.

\item A power-law modification with a more gradual fall-off with distance is
usually associated with higher-dimensional cosmological models:
\beq
U (r) = -\frac{G m_1m_2}{r}\left[ 1 \pm \left( \frac{r_0}{r} \right )^{n}
\right] \, . \label{eq:inverse_square}
\eeq

Both (\ref{eq:yukawa}) and (\ref{eq:inverse_square}) predict a perihelion shift
in the orbit of two bodies which shall be the focus of our discussion in the
present section. For simplicity, we shall restrict our attention to
(\ref{eq:inverse_square}), although the entire discussion carries over to
(\ref{eq:yukawa}) quite simply.

\end{enumerate}

Our experiment proposes to test Newtonian laws of motion by means of two (or
more) test bodies which freely fall within the drag-free environment of a
spin-axis stabilized spacecraft. This spacecraft could be placed either in a
geosynchronous orbit or at the L2 Lagrange point of the Earth--Sun system
(where WMAP is currently deployed). As we proceed to show, such a deployment of
an artificial small-scale planetary system in free fall allows us to probe very
small relative accelerations and test Newton's law on sub-millimeter scales. A
significant advantage of our set-up vis a vis terrestrial experiments is its
conceptual simplicity and its relative freedom from sources of contamination.
(For instance, Casimir forces \cite{casimir}, which can contribute
significantly to the signal on sub-millimeter scales in Cavendish-type
experiments and are difficult to model and subtract, are wholly absent in our
case.)

To illustrate the basic physical principle behind our
experiment, we consider two masses $m_1$ and $m_2$ freely moving in a drag-free
environment and interacting with the potential
\beq
U (r) = -\frac{\alpha}{r} + U_{\rm mod}(r)
\eeq
where $\alpha = G m_1 m_2$, $G$ being the Newton's constant, and $U_{\rm
mod}(r)$ is the new term which modifies the inverse-square law.  We will
consider potentials of the type
\beq
U (r) = -\frac{\alpha}{r}\left[ 1 \pm \left( \frac{r_0}{r} \right )^{n} \right]
\, , \quad r \gg r_0 \, , \quad n \ge 1
\eeq
where $r_0$ is the relevant scale below which gravity becomes essentially
non-Newtonian, and `$\pm$' determines the attractive or repulsive character of
the additional potential. The power $n$ will depend on a particular model; for
instance, $n=2$ in the five-dimensional Randall--Sundrum model \cite{rs99}.

Let us now calculate the perihelion shift in the orbit of these two masses.
Assuming the correction to the potential to be small, we have \cite{mechanics}
\begin{equation} \label{delphi}
\delta \phi = {\partial \over \partial {\cal M}} \left({2 m \over {\cal M}}
\int_0^\pi r^2 U_{\rm mod} (r) d \phi \right) \, ,
\end{equation}
where
\beq
m = \frac{m_1 m_2}{m_1+m_2}~,
\eeq
is the reduced mass, ${\cal M}$ is the angular momentum, and the integral is
taken over the unperturbed orbit
\begin{equation} \label{param}
r = {p \over 1 + e \cos \phi} \, , \quad p = {{\cal M}^2 \over m \alpha} \, , \quad e
= \sqrt{1 + {2 E {\cal M}^2 \over m \alpha^2}} \, ,
\end{equation}
where $e$ is the eccentricity, and $E < 0$ is the total energy.  The partial
derivative with respect to ${\cal M}$ in (\ref{delphi}) is calculated under
the assumption of constant $E$.

In general, the integral in (\ref{delphi}) can be evaluated numerically.  But
in the simplifying case $e \ll 1$, the result can also be obtained analytically
as follows:
\begin{eqnarray}
\int_0^\pi r^2 U_{\rm mod}(r) d \phi = {} \mp \alpha r_0^{n} \int_0^\pi {d \phi
\over r^{n-1}} &=& {} \mp \alpha r_0^{n} p^{1 -n} \int_0^\pi (1 + e \cos
\phi)^{n - 1} d \phi \nonumber \\ &\approx& {} \mp \pi \alpha r_0^{n} p^{1 - n}
\, , \label{intmod}
\end{eqnarray}
where the approximation assumes $e = 0$, and is exact in the case $n = 2$.
 Substituting
 (\ref{intmod}) into (\ref{delphi}), using (\ref{param}), and taking into
 account the relation
 \begin{equation} \label{mom}
 {\cal M} = \sqrt{\alpha m r} \, ,
 \end{equation}
 valid for an (almost) circular orbit, we obtain the contribution to perihelion
 shift from $U_{\rm mod}(r)$:
 \begin{equation}
 \delta \phi = {} \pm  2 \pi (2n - 1) \left(r_0 \over r \right)^{n} \, .
 \label{eq:shift}
 \end{equation}
Notice that the perihelion shift depends only upon the new parameter $n$ (which
is related to the particular theory of modified gravity) and upon the ratio
$r_0/r$, where $r$ is the radius of the orbit and $r_0$ the length scale below
which departures from the $1/r^2$ force-law occur. From (\ref{eq:shift}) we
notice that in order to make $\delta \phi$ large we should (for a fixed $n$)
try and make the radius of the orbit as small as technologically feasible.

A small radius of the orbit is also advantageous from another perspective.
Recall that the orbital period of a test body around a more massive central mass
$M$ is
\beq
\tau = \frac{2\pi r}{v} = \frac{2\pi r^{3/2}}{\sqrt{G M}} \, ,
\eeq
$r$ being the radius of the (almost) circular orbit. Relative to Earth's motion
around the Sun, we have
\beq\label{eq:sun}
\frac{\tau_{\rm orbit}}{\tau_{\rm Earth~ orbit}} = \left (\frac{r}{\rm 1
~au}\right )^{3/2} \left (\frac{M_\odot}{M}\right )^{1/2} = 0.02428 \times
\left (\frac{r}{\rm 1 ~m}\right )^{3/2} \left (\frac{\rm 1 ~kg}{M}\right
)^{1/2} \, ,
\eeq
where $1\,{\rm au} = 1.5\times 10^{11}$~m is the astronomical unit, and
$M_\odot = 1.99\times 10^{30}$~kg is the Solar mass. The mass of a tungsten
sphere ($\rho = 19.6$~g/cm$^3$) of four-centimeter radius is $M = 5254$~g;
hence, if our test body orbits it at a distance of $r = 10$~cm, we get
\beq
\frac{\tau_{\rm Earth~ orbit}}{\tau_{\rm orbit}} = 2985~,
\eeq
i.e., our test mass will make almost $3000$ revolutions per year --- a
significant number\,!

Since the perihelion shift is an additive quantity, the {\em total\/}
perihelion shift in our space-based two-body system in a single year becomes
\beq
\delta\phi_{\rm one~ year} = {\cal N} \delta \phi \, ,
\eeq
where ${\cal N} = \tau_{\rm Earth~ orbit}/\tau_{\rm orbit}$. Substituting from
(\ref{eq:shift}) and (\ref{eq:sun}), we find
\beq
\delta\phi_{\rm one~ year} = \pm 2 \pi (2n - 1) \left (\frac{\rm 1
~au}{r}\right )^{3/2} \left (\frac{M}{M_\odot}\right )^{1/2} \left(r_0 \over r
\right)^{n} \, , \label{eq:total_shift}
\eeq
Since
\beq
\delta\phi_{\rm one~ year} ~\propto~ \left (\frac{M}{r^{3+2n}}\right )^{1/2}~,
\eeq
a larger perihelion shift is obtained by: (i)~making the central mass as large
as possible and (ii)~by simultaneously shrinking the radius of the orbit. (Of
course, in practice, (i) and (ii) work in opposite directions so a judicial
consideration needs to be applied to make an optimum choice.)

Consider next the extra-dimensional scenario due to Randall and Sundrum
\cite{rs99}, in which Newton's law becomes five-dimensional on sub-micron
scales and is modified by the correction $U_{\rm mod} \propto 1/r^3$ on scales
larger than a micron. Substituting $r_0 = 10^{-4}$~cm and $n = 2$ in
(\ref{eq:total_shift}), we obtain
\beq
\delta\phi_{\rm one~ year} \simeq 1 \ \mbox{arc sec} \, .
\eeq
We therefore find that if Newton's law becomes five dimensional on sub-micron
scales, then the cumulative perihelion shift in our space-based two-body system
is about an arc second per year, which is a reasonable quantity to try and
measure.

It is interesting to note that the accelerations experienced by our miniature
planetary system are typically very small. For instance, a test body at a
distance of one meter from the central mass $M$ will experience an acceleration
of only $3\times 10^{-8}$~cm/s$^2$, which is close to the MOND value\,! So,
departures from the Newtonian inertia law should be verifiable for a MOND-type
theory using our space-borne planetary system. In this case, one places two (or
more) bodies in orbit at different radii around $M$. Clearly, the closer body
will probe departures from the inverse-square law whereas the more distant ones
will probe MOND.\footnote{Of central importance here is the question of whether
MOND works in regions of small relative or absolute accelerations. If the
latter is the case, then the much larger acceleration (relative to the MOND
value) experienced by the spacecraft due to its vicinity to the Sun will make
extra-terrestrial tests of MOND quite useless. On the other hand, if the local
acceleration holds the key to the MOND phenomenon, then experiments like the
one proposed in this paper should be able to test this hypothesis. (See
\cite{mond_recent} for recent attempts to provide a theoretical foundation to
MOND and \cite{sanders} for an excellent review.)}

\subsection{Spacecraft Design --- some preliminary ideas}

The following points need to be noted in connection with the design of our
space experiment:

\begin{enumerate}

\item The payload will include, in addition to the test masses, a tracking
camera which will monitor the motion of the miniature planetary system floating
within the spacecraft.\footnote{Our experimental setup has been adapted from an
earlier suggestion for a space experiment \cite{nobili,see}, in which the idea
of a space-based mini-planetary system was advocated with the purpose of
obtaining accurate measurements of Newton's gravitational constant $G$.}

\item Since the masses involved are to be in free fall, it is essential that
all non-gravitational forces are minimized. For this purpose, the spacecraft
will play the role of a Faraday cage and screen the experiment from any
external electric field as well as cosmic rays. In addition, any residual gas
present within the spacecraft can easily be released by means of a small
opening.

\item Since it is essential that the test bodies be allowed to execute their
motion in a drag-free environment, one must account for all non-gravitational
forces which could act on the spacecraft. One of the main perturbations in
outer space is the solar radiation pressure. If the spacecraft is placed in a
geosynchronous orbit then, for an area-to-mass ratio of $0.26$~cm$^2$/g, the
perturbing acceleration generated by the solar wind is $\sim 10^{-5}$~cm/s$^2$,
which is somewhat larger than the relative acceleration between the test masses
in our experiment. Therefore our drag-free spacecraft should be designed to
compensate for this non-gravitational acceleration perhaps by activating jets
which ensure that the spacecraft follows the free-fall motion of the miniature
planetary system. The influence of solar wind can also be minimized by placing
the spacecraft near the L2 Lagrange point which is shielded from the Sun by the
Earth's shadow.

Also note that it is possible for charged particles (mostly protons) to penetrate
the spacecraft thereby creating strong electrostatic fields which could disturb
the experiment. Of the three sources of such contamination: cosmic rays, solar
flares and the Van Allen belts, the latter are the most dangerous \cite{see}.
However because the belts are associated with the Earth, one can avoid this
effect by having the spacecraft at the L2 point instead of in a geosynchronous
orbit. This could also decrease the solar flare component, although for complete
safety, the experiment should be `switched off' during (rare) periods of intense solar
activity.

\item An important non-rigid perturbation source is the fuel whose gravity
gradient effects on the experiment need to be modelled. Clearly, in order to
minimize `force-contamination', the fuel tank must be placed as far away from
the experiment as possible. Keeping in mind that the radius of the experiment
is $ < 1$~m, we suggest (following Nobili {\em et al\@.} \cite{nobili}) that
the spacecraft be cylindrical: 3.5 meters in height with a base diameter of 3
meters and a mass of 400~kg. Furthermore, the spin axis of the spacecraft is
stabilised with a spin period of 60 seconds.\footnote{Spinning the spacecraft
is important because it averages forces which are body-based in the spin plane
which is also the orbital plane of the planets. Consequently only the zonal
harmonics of the gravitational field of the spacecraft are relevant in this
case \cite{nobili,powel_deBra}.} It may be noted that, in order for an
experiment to last an interval of time $\Delta T$, the mass of liquid
propellant fuel required is \cite{nobili}
\beq
\left (\frac{\Delta M}{1~ {\rm kg}}\right ) = 5.2\times 10^{-3} \left
(\frac{\Delta T} {1 {\rm ~day}}\right ) \, .
\eeq
Since our experiment is likely to be operational for a two-year duration (the
longer the better\,!), its fuel requirement is $\Delta M \simeq 4$~kg.

Also note that tidal effects on the spacecraft due to the gravitational
field of the environment could be a potential source of
perihelion shift of our planetary system and need to be properly
understood and incorporated into the analysis.

\item Some departure from spherical symmetry is likely to occur for the
`planets' due to purely technological reasons. This effect needs to be included
in the analysis perhaps by expanding the gravitational field in spherical
harmonics --- a standard exercise in satellite geodesy.

\item A mass-release mechanism which will release the `planets' and place them
in an elliptical orbit is essential. Since the spacecraft is spinning much more
rapidly than the orbital period of the planets, the mechanism which effects
mass-release must do so gently, so that planets are injected with a relative
velocity which is smaller than their escape velocity (with respect to each
other). Some ideas for this have been put forth in \cite{nobili} and the reader
is referred to these papers for more details. Proper account must also be taken
of the planets' spin in order to preclude the occurrence of a spin-orbital
resonance. A central role in this experiment will be an accurate measurement of
the semi-major axis required for determining the perihelion shift.

\item Even with an excellent mass-release mechanism, it is still possible that
the two planets will not be placed into the proper orbital plane. (To ensure
proper tracking it is essential that the inclination of the orbital plane be
close to the equatorial plane of the spacecraft which will also be the camera's
focal plane.) In this case, orbit correction devices, such as light sources
capable of exerting a gentle pressure on the `planets', need to be incorporated
into the design of the spacecraft. (An acceleration of $\sim 10^{-8}$~cm/s$^2$
can be induced on our `planet' by means of a 10~W source of light concentrated
in a 10 degree cone. So one might expect that intermittent bursts of light
could be used to gently change the inclination of the orbit.)

\item Note that our experimental set-up requires a prior determination of $M$
which can easily be carried out in a laboratory on Earth. Thus all essential
experimental inputs can be determined to high accuracy terrestrially while the
motion of our miniature planetary system is easy to predict theoretically and
can therefore be compared with observational measurements made within the
spacecraft.

\item Finally, note that at least some of the technology required for APSIS is
already being developed in connection with the LISA Pathfinder (LP) mission,
which envisages two test objects in gravitational free fall in a drag-free
environment, and in the Gravity Probe B (GP-B) gyroscope experiment. The LP
test objects are shielded from non-gravitational forces and will be discharged
at regular intervals using fibre-coupled UV lamps \cite{LP}. A gentle release
mechanism for the test masses is crucial for the LISA Pathfinder, in which one
hopes to achieve a release speed of less than $5~\mu$m/s (18~mm per hour) and
thereafter measure the position of the test masses with respect to the
spacecraft (or each other) to an accuracy of $10^{-12}$~m.  The designers of
the GP-B experiment have achieved sphericity of the gyroscope spheres (made of
homogeneous fused quartz of 3.81 centimeter in diameter) less than 40 atomic
layers from perfect.\footnote{See http://einstein.stanford.edu}

\end{enumerate}

\section{Conclusions}

It is well known that, within Newtonian mechanics, closed particle orbits occur
only for two types of central force fields \cite{mechanics}, namely, the
inverse-square law $F \propto r^{-2}$ and the harmonic oscillator $F \propto
r$. A force law which deviates on small scales from $1/r^2$ is therefore
expected to give rise to particle orbits which show small perihelion shifts.
This effect is additive in nature since the perihelion shift for successive
orbits is added to give the {\em total\/} shift in the major (minor) axis of a
planet within a stipulated time scale (say, a year). Since smaller orbits also
have shorter periods of rotation, it follows that the net magnitude of this
effect is larger for planets with smaller orbits. Keeping this in mind, we are
of the view that a miniature planetary system placed in free fall within a
spacecraft located at the L2 Lagrange point of the Earth--Sun system, provides
an ideal testing ground for possible departures of Newtonian gravity from the
familiar inverse-square law. Since many higher-dimensional cosmological models
(and several models of modified gravity) predict such departures on small
scales, our Artificial Planetary System In Space (APSIS) could provide a space
laboratory with which to test such theories. The small accelerations prevalent
in such systems $(\sim 10^{-8}$~cm/s$^2)$ may also be useful for probing the
MOND hypothesis. In our experimental setup, one (or more) test bodies (planets)
orbit a more massive central $\sim 5$~kg mass at a distance of $\sim 10$~cm.
The perihelion shift for this configuration is $\sim 1$~arc second/year if the
spatial scale for the modification of gravity law is of the order of a micron
(1~$\mu$m).

\section*{Acknowledgments}

The authors acknowledge interesting discussions with E.~Fischbach,
C.~Laemmerzahl, J.~Lasue, A.~N.~Ramprakash and R.~H.~Sanders. The authors also
acknowledge support from the Indo-Ukrainian program of cooperation in science
and technology sponsored by the Department of Science and Technology of India
and Ministry of Education and Science of Ukraine.


\begin{thebibliography}{99}

\small

\bibitem{sn}
D.~N.~Spergel {\em et al.}, Astrophys. J. Suppl. {\bf 148} (2003) 175
[arXiv:astro-ph/0302209]; \\
A.~G.~Riess {\em et al.}, Astrophys. J. {\bf 607} (2004) 665
[arXiv:astro-ph/0402512].

\bibitem{eisenstein05}
D.~J.~Eisenstein \etal, \apj {\bf 633} (2005) 560 [arXiv:astro-ph/0501171].

\bibitem{astier05}
P.~Astier \etal, Astron. Astrophys. {\bf 447} (2006) 31
[arXiv:astro-ph/0510447].

\bibitem{sahni_greece}
V.~Sahni, Lect. Notes Phys. {\bf 653} (2004) 141 [arXiv:astro-ph/0403324].

\bibitem{abazajian06}
K.~Abazajian, Phys. Rev. D {\bf 73} (2006) 063513 [arXiv:astro-ph/0512631].

\bibitem{sahni_wang}
V.~Sahni and L.~Wang, \pd {\bf 62} (2000) 103517 [arXiv:astro-ph/9910097].

\bibitem{um00}
L.~A.~Ure\~{n}a-L\'{o}pez and T.~Matos, \pd {\bf 62} (2000) 081302
[arXiv:astro-ph/0003364].

\bibitem{hu_barkana-gruz}
W.~Hu, R.~Barkana, and A.~Gruzinov, \prl {\bf 85} (2000) 1158
[arXiv:astro-ph/0003365].

\bibitem{milgrom}
M.~Milgrom, \apj {\bf 270} (1983) 365; {\bf 270} (1983) 371; {\bf 270} (1983)
384.

\bibitem{sanders}
R.~H.~Sanders, {\em Modified gravity without dark matter}, lecture given at
Third Aegean Summer School, The Invisible Universe: Dark Matter and Dark
Energy, arXiv:astro-ph/0601431.

\bibitem{DE_review1}
V. Sahni and A.~A.~Starobinsky, Int. J. Mod. Phys. {\bf D9}, (2000) 373
[arXiv:astro-ph/9904398];
Int. J. Mod. Phys. {\bf D15} (2006) 2105
[arXiv:astro-ph/0610026];
S~M. Carroll, Living Rev.Rel. {\bf 4}, 1 (2001) [{\tt astro-ph/0004075}];
P.~J~E.~Peebles and B.~Ratra, Rev.Mod.Phys. {\bf 75}, (2003) 559
[arXiv:astro-ph/0207347];
T. Padmanabhan, Phys. Rep. {\bf 380}, (2003) 235
[arXiv:hep-th/0212290];
E. J. Copeland, M. Sami and S. Tsujikawa,
Int. J. Mod. Phys. {\bf D15}, (2006) 1753 [arXiv:hep-th/0603057].

\bibitem{brane}
G. Dvali, G. Gabadadze and  M. Porrati, 2000, \plb {\bf 485}, 208
[arXiv:hep-th/0005016];
V. Sahni and Yu. Shtanov, 2003, JCAP {\bf 0311}, 014 [arXiv:astro-ph/0202346];
V. Sahni, {\em Cosmological Surprises from Braneworld models of Dark Energy},
arXiv:astro-ph/0502032.

\bibitem{DE_review2}
S.~Nojiri and S.~D.~Odintsov, Int. J. Geom. Meth. Mod. Phys. {\bf 4} (2007) 115
[arXiv:hep-th/0601213]; R.~P.~Woodard, 2006, {\em Avoiding Dark Energy with 1/R
Modifications of Gravity} arXiv:astro-ph/0601672.

\bibitem{alternatives1}
  M.-N.~C{\'e}l{\'e}rier,
  Astron. Astrophys.  {\bf 353} (2000) 63 [arXiv:astro-ph/9907206];
  R.~K.~Barrett and C.~A.~Clarkson,
  Class. Quantum Grav. {\bf 17} (2000) 5047 [arXiv:astro-ph/9911235];
  K.~Tomita,
  Mon. Not. Roy. Astron. Soc. {\bf 326} (2001) 287 [arXiv:astro-ph/0011484];
  K.~Tomita,
  Prog. Theor. Phys. {\bf 106} (2001) 929 [arXiv:astro-ph/0104141];
  H.~Iguchi, T.~Nakamura and K.~I.~Nakao,
  Prog. Theor. Phys.  {\bf 108} (2002) 809 [arXiv:astro-ph/0112419];
  J.~W.~Moffat,
  JCAP {\bf 0605} (2006) 001;
  H.~Alnes, M.~Amarzguioui and {\O}.~Gr{\o}n,
  Phys. Rev. D {\bf 73} (2006) 083519 [arXiv:astro-ph/0512006];
  R.~Mansouri, {\em Structured FRW universe leads to acceleration: a non-perturbative approach},
  arXiv:astro-ph/0512605;
  C.~H.~Chuang, J.~A.~Gu and W.~Y.~Hwang,
  {\em Inhomogeneity-Induced Cosmic Acceleration in a Dust
  Universe}, arXiv:astro-ph/0512651;
  R.~A.~Vanderveld, E.~E.~Flanagan and I.~Wasserman,
  Phys. Rev. D {\bf 74} (2006) 023506 [arXiv:astro-ph/0602476];
  D.~Garfinkle,
  Class. Quant. Grav.  {\bf 23} (2006) 4811 [arXiv:gr-qc/0605088];
  T.~Biswas, R.~Mansouri and A.~Notari,
  {\em Nonlinear Structure Formation  and ``Apparent'' Acceleration: an
  Investigation},  arXiv:astro-ph/0606703;
  D.~J.~H.~Chung and A.~E.~Romano,
  Phys. Rev. D {\bf 74} (2006) 103507 [arXiv:astro-ph/0608403];
  K.~Enqvist and T.~Mattsson, JCAP {\bf 0702} (2007) 019 [arXiv:astro-ph/0609120];
  H.~Alnes and M.~Amarzguioui,
  Phys. Rev. D {\bf 75} (2007) 023506 [arXiv:astro-ph/0610331];
  %\bibitem{Celerier:2007}
  M.-N.~C{\'e}l{\'e}rier, New Advances in Physics {\bf 1} (2007) 29 [arXiv:astro-ph/0702416];
  R.~R. Caldwell and A. Stebbins,
  {\em A Test of the Copernican Principle}, arXiv:0711.3459.

\bibitem{alternatives2}
F.~I.~Cooperstock and S.~Tieu, Mod. Phys. Lett. A {\bf 21} (2006) 2133;
F.~I.~Cooperstock and S.~Tieu, Int. J. Mod. Phys. A {\bf 22} (2007) 2293
[arXiv:astro-ph/0610370]; H.~Balasin and D.~Grumiller, {\em Significant
reduction of galactic dark matter by general relativity},
arXiv:astro-ph/0602519; M.~Korzy\'{n}ski, {\em Singular disk of matter in the
Cooperstock--Tieu galaxy model}, arXiv:astro-ph/0508377; D.~Vogt and
P.~S.~Letelier, {\em Presence of exotic matter in the Cooperstock and Tieu
galaxy model}, arXiv:astro-ph/0510750; D.~L.~Wiltshire, New J. Phys. {\bf 9}
(2007) 377 [arXiv:gr-qc/0702082]; D.~L.~Wiltshire, {\em Exact solution to the
averaging problem in cosmology}, arXiv:0709.0732; T.~Buchert, {\em Dark Energy
from Structure --- A Status Report}, arXiv:0707.2153.

\bibitem{pioneer}
J.~D.~Anderson {\em et al.}, \prl {\bf 81} (1998) 2858 [arXiv:gr-qc/9808081].

\bibitem{pioneer1}
S.~G.~Turyshev, M.~M.~Nieto, and J.~D.~Anderson, EAS Publ. Ser. {\bf 20} (2006)
243 [arXiv:gr-qc/0510081].

\bibitem{nasa}
S.~G.~Turyshev, M.~M.~Nieto, and J.~D.~Anderson, {\em A Route to Understanding
of the Pioneer Anomaly\/}, arXiv:gr-qc/0503021.

\bibitem{kk}
T.~Appelquist, A.~Chodos, and P.~G.~O.~Freund (editors), {\sl Modern
Kaluza-Klein Theories}, Addison-Wesley Publishing Co. (1987).

\bibitem{antoniadis}
I.~Antoniadis, Phys. Lett. B {\bf 246} (1990) 377.

\bibitem{dvali}
N.~Arkani-Hamed,  S.~Dimopoulos, and G.~Dvali, Phys. Lett. B {\bf 429}
(1998) 263 [arXiv:hep-ph/9803315]; \\
I.~Antoniadis, N.~Arkani-Hamed, S.~Dimopolous, and G.~Dvali, Phys. Lett. B {\bf
436} (1998) 257 [arXiv:hep-ph/9804398].

\bibitem{ss02a}
V.~Sahni and Yu.~Shtanov, Int. J. Mod. Phys. D {\bf 11} (2002) 1515
[arXiv:gr-qc/0205111].

\bibitem{hoyle01}
C.~D.~Hoyle {\em et al.} \prl {\bf 86} (2001) 1418 [arXiv:hep-ph/0011014]; \\
C.~D.~Hoyle, D.~J.~Kapner, B.~R.~Heckel, E.~G.~Adelberger, J.~H.~Gundlach,
U.~Schmidt, and H.~E.~Swanson, \pd {\bf 70} (2004) 042004
[arXiv:hep-ph/0405262].

\bibitem{rs99}
L.~Randall and R.~Sundrum, Phys.\@ Rev.\@ Lett.\@ {\bf 83} (1999) 3370
[arXiv:hep-ph/9905221]; \\
L.~Randall and R.~Sundrum, Phys.\@ Rev.\@ Lett.\@ {\bf 83} (1999) 4690
[arXiv:hep-th/9906064].

\bibitem{rubsh}
V.~A.~Rubakov and M.~E.~Shaposhnikov, \plb {\bf 152} (1983) 136.

\bibitem{akama}
M.~Akama, Prog. Theor. Phys. {\bf 78} (1987) 184.

\bibitem{rub_uspekhi}
V.~A.~Rubakov, Phys. Usp. {\bf 44} (2001) 871 [Usp. Fiz. Nauk {\bf 171} (2001)
913] [arXiv:hep-ph/0104152].

\bibitem{callin_ravndall}
P.~Callin and F.~Ravndal, Phys. Rev. D {\bf 70} (2004) 104009
[arXiv:hep-ph/0403302].

\bibitem{hwitten}
P.~Ho\v{r}ava and E.~Witten, \nucp B {\bf 460} (1996) 506 [arXiv:hep-th/9510209]; \\
P.~Ho\v{r}ava and E.~Witten, \nucp B {\bf 475} (1996) 94
[arXiv:hep-th/9603142].

\bibitem{maartens}
R.~Maartens, Living Rev. Relativity {\bf 7}(2004) 7 [arXiv:gr-qc/0312059].

\bibitem{bronnikov06}
K.~A.~Bronnikov, S.~A.~Kononogov, and V.~N.~Melnikov, Gen. Rel. Grav. {\bf 38}
(2006) 1215 [arXiv:gr-qc/0601114].

\bibitem{navarro05}
I.~Navarro and K.~Van Acoleyen, JCAP {\bf 0609} (2006) 006
[arXiv:gr-qc/0512109].

\bibitem{sahni_kyoto}
V.~Sahni, {\em Cosmological surprises from Braneworld models of dark energy},
in: Proceedings of the 14th Workshop on General Relativity and Gravitation
(JGRG14), Kyoto University, Japan 29 November -- 3 December, 2004. Edited by
W.~Hikida, M.~Sasaki, T.~Tanaka and T.~Nakamura, pp.~95--115
[arXiv:astro-ph/0502032].

\bibitem{modified-grav}
S.~Nojiri and S.~D.~Odintsov, Int. J. Geom. Meth. Mod. Phys. {\bf 4}
(2007) 115 [arXiv:hep-th/0601213]; \\
R.~P.~Woodard, {\em Avoiding Dark Energy with 1/R Modifications of Gravity},
arXiv:astro-ph/0601672.

\bibitem{casimir}
H.~B.~G.~Casimir, Proc. K. Ned. Akad. Wet. {\bf 51} (1948) 635; \\
E.~G.~Adelberger, B.~R.~Heckel, and A.~E.~Nelson, Ann. Rev. Nucl. Part. Sci.
{\bf 53} (2003) 77 [arXiv:hep-ph/0307284]; \\
G.~L.~Klimchitskaya, R.~S.~Decca, E.~Fischbach, D.~E.~Krause, D.~Lypez, and
V.~M.~Mostepanenko, Int. J. Mod. Phys. A {\bf 20} (2005) 2205
[arXiv:quant-ph/0506120].

\bibitem{mechanics}
L.~D.~Landau and E.~M.~Lifshitz, {\sl Mechanics}, Butterworth-Heinemann, 1996.

\bibitem{mond_recent}
M.~Milgrom, Annals Phys. {\bf 229} (1994) 384 [arXiv:astro-ph/9303012]; \\
M.~Milgrom, Phys. Lett. A {\bf 253} (1999) 273 [arXiv:astro-ph/9805346]; \\
J.~D.~Bekenstein, \pd {\bf 70} (2004) 083509 [arXiv:astro-ph/0403694]; \\
J.~D.~Bekenstein and R.~H.~Sanders, {\em A Primer to Relativistic MOND Theory},
arXiv:astro-ph/0509519; \\
M.~Milgrom, {\em MOND as Modified Inertia}, arXiv:astro-ph/0510117; \\
J.~Bekenstein and J.~Magueijo, Phys. Rev. D {\bf 73} (2006) 103513
[arXiv:astro-ph/0602266].

\bibitem{nobili}
A.~M.~Nobili, A.~Milani, and P.~Farinella, Phys. Lett. A {\bf 120}
(1987) 437; \\
A.~M.~Nobili, A.~Milani, and P.~Farinella, Astron. J. {\bf 95} (1988) 576; \\
A.~M.~Nobili, A.~Milani, E.~Polacco, I.~W.~Roxburgh, F.~Barlier, K.~Aksnes,
C.~W.~F.~Everitt, P.~Farinella, L.~Anselmo, and Y.~Boudon, {\em The Newton
Mission - A proposed manmade planetary system in space to measure the
gravitational constant}, ESA Journal {\bf 14} (1990) 389.

\bibitem{see}
A.~J.~Sanders and W.~E.~Deeds, Phys. Rev. D {\bf 46} (1992) 480; \\
A.~J.~Sanders {\em et al.}, Meas. Sci. Technol. {\bf 10} (1999) 514; \\
A.~J.~Sanders {\em et al.}, Class. Quantum Grav. {\bf 17} (2000) 2331; \\
A.~D.~Alexeev {\em et al.}, Grav. Cosmol. {\bf 5} (1999) 67
[arXiv:gr-qc/0002088].

\bibitem{powel_deBra}
J.~D.~Powell and D.~B.~de~Bra, Revue RAIRO Avril-J-1, pp. 88--99 (1974).

\bibitem{LP}
D.~Bortoluzzi, \etal, Class. Quantum Grav. {\bf 21} (2004) S573
[arXiv:gr-qc/0402020]; \\
http://sci.esa.int/science-e/www/object/index.cfm?fobjectid=35589

\end{thebibliography}
\end{document}